\documentclass[prl,twocolumn,superscriptaddress,floatfix]{revtex4-2}%
\usepackage{graphicx}
\usepackage{bm, amsmath, amssymb}
\usepackage{color}
\usepackage{soul}
\usepackage{amstext}
\usepackage{amsmath}
\usepackage{amsfonts}
\usepackage{amssymb}
\usepackage[T1]{fontenc}%
\usepackage{hyperref}

\begin{document}
\title{Scalable Robust Quantum Control for Semiconductor Spin Qubits with Always-on Couplings}

\author{Yong-Ju Hai}
\affiliation{Shenzhen International Quantum Academy (SIQA), Futian District, Shenzhen, P. R. China}

\author{Shihang Zhang}
\author{Haoyu Guan}
\affiliation{Shenzhen Institute for Quantum Science and Engineering (SIQSE), Southern University of Science and Technology, Shenzhen, P. R. China}
\author{Peihao Huang}
\author{Yu He}
\author{Xiu-Hao Deng}
\email{ dengxh@sustech.edu.cn }
\affiliation{Shenzhen International Quantum Academy (SIQA), Futian District, Shenzhen, P. R. China}
\affiliation{Shenzhen Institute for Quantum Science and Engineering (SIQSE), Southern University of Science and Technology, Shenzhen, P. R. China}

\begin{abstract}
We demonstrate a robust quantum control framework that enables high-fidelity gate operations in semiconductor spin qubit systems with always-on couplings. Always-on interactions between qubits pose a fundamental challenge for quantum processors by inducing correlated errors that can trigger chaotic dynamics. Our approach suppresses both static coupling noise and time-dependent crosstalk without requiring high on/off ratio tunable couplers. Significantly, these pulses also prevent the emergence of chaotic entanglement growth in deep quantum circuits, preserving coherence in large multi-qubit systems. By relaxing hardware constraints on coupling control, our method provides a practical path toward scaling semiconductor quantum processors within existing fabrication capabilities, with particular relevance for silicon spin qubit architectures where high-contrast coupling modulation remains challenging.
\end{abstract}

\maketitle
Semiconductor spin qubits represent a promising platform for large-scale quantum computation due to their long coherence times~\cite{burkard2023semiconductor, RevModPhys.85.961, xue2022quantum, noiri2022fast, mkadzik2022precision, philips2022universal, huang2019fidelity}, small footprint, and compatibility with standard semiconductor manufacturing processes. However, a critical challenge in scaling these systems arises from always-on couplings between neighboring qubits, which induce qubit crosstalk~\cite{Tanttu2024-nu, zhao2022quantum}, spectrum broadening~\cite{yoneda2021coherent,deng2021correcting}, and correlated errors~\cite{Yoneda2023-wy, PhysRevB.101.235133, yi2024robust,deng2021correcting} that undermine quantum error correction~\cite{hwang2001correlated,clader2021impact,bombin2016resilience}. Most concerning, they trigger unwanted entanglement growth across the system that ultimately leads to chaotic dynamics~\cite{berke2022transmon,borner2023classical}. The problem is particularly acute in silicon spin qubit architectures, where high-contrast modulation of inter-qubit coupling strengths remains technically challenging~\cite{xue2022quantum,noiri2022fast, miao2023overcoming,kono2024mechanically}, limiting the effectiveness of conventional decoupling protocols that rely on rapid switching of interaction strengths.

Previous approaches to managing always-on couplings have focused on hardware-based solutions, including the development of tunable couplers with high on/off ratios or an engineered disorder to suppress error propagation~\cite{xue2022quantum, Mills2022, Huang2024-fz, miao2023overcoming,kono2024mechanically}. However, these strategies impose substantial fabrication demands and require precise control over a large parameter space. Alternatively, dynamical decoupling techniques~\cite{khodjasteh2007performance} can mitigate certain types of noise but typically address only specific error channels and may themselves introduce additional crosstalk~\cite{yi2024robust} when implemented across multiple qubits simultaneously. These limitations highlight the need for a comprehensive control framework that can operate within the constraints of existing hardware while addressing the full spectrum of coupling-induced errors.
 
In this Letter, we demonstrate a robust quantum control framework that enables high-fidelity robust operations in semiconductor spin qubit systems with always-on couplings. Our approach suppresses both static coupling noise and time-dependent crosstalk, without requiring modifications to the hardware architecture. These pulses not only maintain gate fidelity but also prevent chaotic entanglement growth through deep quantum circuits. By relaxing hardware constraints on coupling control, our method provides a viable path toward scaling semiconductor quantum processors within existing fabrication capabilities.

\begin{figure}[tb]
	\centering
	\includegraphics[width=\columnwidth]{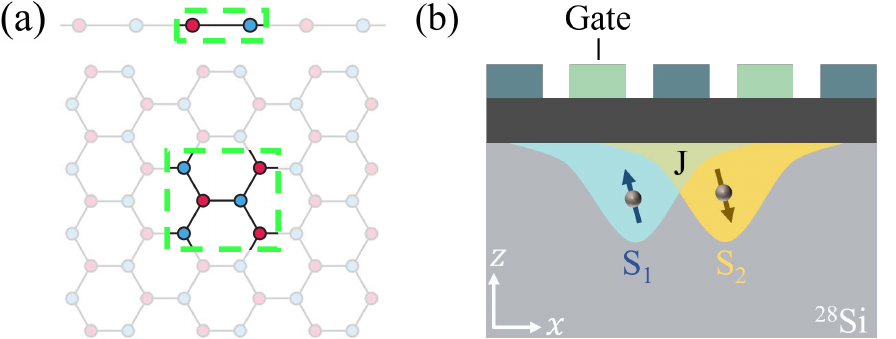}
	\caption{ (a) 1D and 2D multi-qubit architectures. The highlighted regions with green square dashed lines indicate the fundamental unit in the always-on coupling system.
		(b) The schematic diagram of gate-defined double quantum dots with always-on coupling as an example of the simplest unit in (a). The always-on exchange coupling is represented by the overlap of the wavefunctions of the two electrons (blue and yellow area).}
	\label{Fig_1_architecture}
\end{figure}

\paragraph{Errors in Multi-Qubit System.-}We consider multi-qubit systems with always-on couplings, focusing on one-dimensional chain and two-dimensional honeycomb lattices, as shown in Fig.~\ref{Fig_1_architecture}(a). When implementing quantum gates, we must consider the influence of neighboring qubits, namely spectators that are weakly coupled to targeted qubits~\cite{singh2023mid} and intrusions with a stronger coupling that significantly affects system dynamics~\cite{deng2021correcting}. These couplings hybridize the computational basis states and introduce various crosstalk effects that must be explicitly modeled.

Consider a quantum system with $d$ target qubits (on which we implement quantum gates) coupled to $m$ neighboring qubits (spectators or intruders). The total Hamiltonian can be expressed as $H_{\text{tot}} = H_0 + V(t)$, where $H_0$ is the native Hamiltonian and $V(t)$ represents the control field. In the eigenbasis of $H_0$, this Hamiltonian adopts a block structure:
\begin{equation}
\tilde{H}^{m;d} = \bigoplus_{i=1}^{2^m} \tilde{H}_i + \bigoplus_{i,j=1}^{2^m} \tilde{V}_{ij}(t).
\label{Eq_GeneralH}
\end{equation}
Here, the superscript $m;d$ denotes a system with $m$ neighboring qubits affecting $d$ target qubits. The Hamiltonian decomposes into $2^m\times2^m$ blocks corresponding to different configurations of the neighboring qubits. For example, in a system with two neighboring qubits and one target qubit, $\text{span}\{\left|\uparrow\uparrow;\uparrow\right\rangle, \left|\uparrow\uparrow;\downarrow\right\rangle\}$ represents the $i=1$ diagonal subspace where both neighboring qubits are in the $\left|\uparrow\right\rangle$ state.

Without loss of generality, we designate the $i=1$ subspace as our reference and define the reference Hamiltonian $\tilde{H}_{\text{r}}=\tilde{H}_1+\tilde{V}_{11}$. The difference between the reference and the original Hamiltonian is identified as the noise Hamiltonian:
\begin{equation}
\tilde{H}_{\text{n}} = \tilde{H}^{m;d}-\tilde{H}_{\text{r}}=\bigoplus_{i=2}^{2^m} (\Delta\tilde{H}_i + \Delta \tilde{V}_{ii}) + \bigoplus_{i,j=1,i\neq j}^{2^m} \tilde{V}_{ij}.
\end{equation}
Here, $\Delta\tilde{H}_i=\tilde{H}_i-\tilde{H}_1$ represents frequency shifts experienced by target qubits due to different configurations of neighboring qubits, collectively forming a manifold of $2^m$ distinct energy levels. The terms $\Delta \tilde{V}_{ii}=\tilde{V}_{ii}-\tilde{V}_{11}$ represent variations in control field effects across different neighboring-qubit configurations, while $\tilde{V}_{ij}(t)$ for $i\neq j$ captures crosstalk between different subspaces. These noise terms introduce correlated errors, parasitic operations, and cross-coupling effects that degrade gate fidelities.

To analyze error dynamics, we decompose the total evolution operator as $U = U_0 U_{\text{e}}$, separating the ideal evolution $U_0(t) = \mathcal{T} \exp\{-i \int_0^t d\tau \tilde{H}_{\text{r}}(\tau)\}$ from the error evolution $U_{\text{e}}(t) = \mathcal{T} \exp\{-i \int_0^t d\tau \tilde{V}_{\text{I}}(\tau)\}$. Here, $\tilde{V}_{\text{I}}= U_0^{\dagger} \tilde{H}_{\text{n}} U_0$ represents the noise Hamiltonian in the interaction picture. For any operator $K$, we define the super-operator $\mathcal{R}(K(t))=\int_0^t d\tau U_0(\tau)^{\dagger} K(\tau) U_0(\tau)$, which represents the time-integrated effect of $K$ as transformed by the ideal evolution. In the Pauli basis $\{\sigma^d_\nu\}=\{X,Y,Z,I\}^{\otimes d}$, an operator $K=\mathbf{T}\cdot \mathbf{\sigma}^d$ can be conceptualized as a point moving with velocity $\mathbf{T}$ in operator space, and $\mathcal{R}(K(t))=\mathbf{r}\cdot \mathbf{\sigma}^d$ represents the integrated path traced by this point in the Pauli frame~\cite{hai2022universal}.

Assuming the noise terms are small relative to the baseline Hamiltonian, we can expand the error evolution operator to first order:
\begin{equation}
\begin{aligned}
U_{\text{e}}(t) &=\mathcal{T} \exp\{-i \mathcal{R}(\tilde{H}_{\text{n}}(t))\}\\
&=\prod_{j=1}^{2^m}e^{-i\mathcal{R}(\Delta\tilde{H}_j)}\prod_{j=1}^{2^m}e^{-i\mathcal{R}(\Delta\tilde{V}_j)}\prod_{j=1}^{2^m}\prod_{l>j}^{2^m}e^{-i\mathcal{R}(\tilde{V}_{jl})}\\
&\approx I_{md} - i\sum_{\mu} \mathcal{R}(K_\mu)=I_{md} - i\sum_{\mu}\mathbf{r}_\mu(t)\cdot \mathbf{\sigma}^d,
\end{aligned}
\label{Eq_GeneralUe}
\end{equation}
where each noise term $K_\mu$ belongs to the set $\{\Delta\tilde{H}_j,\Delta\tilde{V}_j, \tilde{V}_{jl} \}$. Each error component generates an error curve $\mathbf{r}_\mu(t)$ in operator space, with the vector norm $\|\mathbf{r}_\mu(t)\|$ quantifying the magnitude of the associated error. 
We define the total error distance as a measure of control robustness against all noises
\begin{equation}
D = 2^{-\frac{d}{2}} \sum_\mu \left\|\mathbf{r}_\mu\cdot \mathbf{\sigma}^d\right\|_{\text{F}}
= \sum_\mu \|\mathbf{r}_\mu(t)\|.
\end{equation}
Here $\|\cdot\|_{\text{F}}$ denotes the Frobenius norm. For perfect gate implementation with high precision and robustness, we require $U_0(T) = U_{\text{target}}$ or the noiseless gate fidelity $F \approx 1$ and error distance $D=0$. Therefore, $D$ serves as the error-correcting constraint when designing robust control pulses~\cite{xue2024traversing}.

\paragraph{Coupled Quantum Dot.-}To illustrate our approach, we consider a common example: a pair of coupled gate-defined quantum dots, as shown in Fig.~\ref{Fig_1_architecture}(b). The spin states follow an extended Heisenberg Hamiltonian~\cite{burkard2023semiconductor}
\begin{equation}
  H = \mathbf{B}_1 \cdot \mathbf{S}_1 + \mathbf{B}_2 \cdot \mathbf{S}_2 + J
  \left( \mathbf{S}_1 \cdot \mathbf{S}_2 - \frac{1}{4} \right).
  \label{Eq_Heisenberg_Hamiltonian}
\end{equation}
Here, $\mathbf{S}_j = (X_j, Y_j, Z_j) / 2$ are the spin operators, and $\mathbf{B}_j = (B_{x, j}, B_{y, j}, B_{z, j})$ represents the magnetic field at each qubit. The z-components of the magnetic fields determine the electron spin resonance frequencies, while transverse fields provide qubit control. The exchange coupling $J$ brings the two spins into the coupled basis $\{\left|\uparrow \uparrow \right\rangle, |\tilde{\uparrow \downarrow}\rangle, |\tilde{\downarrow \uparrow}\rangle, \left|\downarrow \downarrow \right\rangle\}$, in which the Hamiltonian without transverse controls is diagonalized as $\tilde{H}_0 = \text{diag} \{ 2 E_z, -
\Delta \tilde{E}_z - J, \Delta \tilde{E}_z - J, - 2 E_z \}$, where $\Delta \tilde{E}_z = \sqrt{J^2 + \Delta E_z^2}$ , $|\tilde{\uparrow \downarrow}\rangle=\cos\theta|\uparrow \downarrow\rangle+\sin\theta| \downarrow\uparrow\rangle$, $ |\tilde{\downarrow \uparrow}\rangle=\sin\theta|\uparrow \downarrow\rangle+\cos\theta| \downarrow\uparrow\rangle$, and $\tan \theta = \frac{J}{\Delta E_z + \Delta \tilde{E}_z}$.

When driving qubit 2 with a transverse field $V(t) = \Omega_2(t)/2 X_2$ and transforming to the rotating frame, we obtain
\begin{equation}
    \tilde{H}^{1;1} = \begin{pmatrix}
        \tilde{H}_1 + \tilde{V}_{11} & \tilde{V}_{12} \\
        \tilde{V}_{21} & \tilde{H}_2 + \tilde{V}_{22}
    \end{pmatrix},
    \label{Eq_2dot_H2_full}
\end{equation}
where the components are given by $\tilde{H}_{1,2}= \text{diag} \{\pm J/4, \mp J/4 \}$, $\tilde{V}_{11} = \tilde{V}_{22} = \frac{\Omega_{2}}{2} X$, $\tilde{V}_{12,21} = \frac{1}{2} e^{\pm i\tilde{E}_z t} \tan \theta \Omega_2 Z$. In the coupled Pauli basis, it is simplified as
\begin{equation}
    \begin{aligned}
        \tilde{H}^{1;1} &= \frac{\Omega_2(t)}{2} IX + \frac{J}{4} ZZ
         + \frac{\tan\theta \Omega_2(t)}{2} \\ 
         & \times
         \Big[\cos(\Delta \tilde{E}_z t) XZ - \sin(\Delta \tilde{E}_z t) YZ \Big],
    \end{aligned}
    \label{Eq_2dot_H_Pauli}.
\end{equation}
This Hamiltonian contains the intended control term ($IX$) along with parasitic terms: always-on $ZZ$ coupling and crosstalk terms ($XZ$, $YZ$). In experimentally relevant regimes where $J \ll \Delta E_z$, we consider $J/\Delta E_z < 0.1$ and get $\tan \theta \approx J/2\Delta E_z$. The magnitudes of crosstalk noise and always-on coupling scale as $|\tan\theta \Omega_2|$ and $J$, respectively.
While crosstalk is often negligible in weak-drive regimes~\cite{russ2018high, gungordu2020robust, kanaar2021single}, it becomes significant under strong-drive conditions needed for fast gate operations - thereby necessitating simultaneous suppression of both error channels~\cite{heinz2024analysis, mkadzik2022precision} for high-fidelity operations in multi-qubit systems.

\begin{figure}[t] 
\centering
\includegraphics[width=\columnwidth]{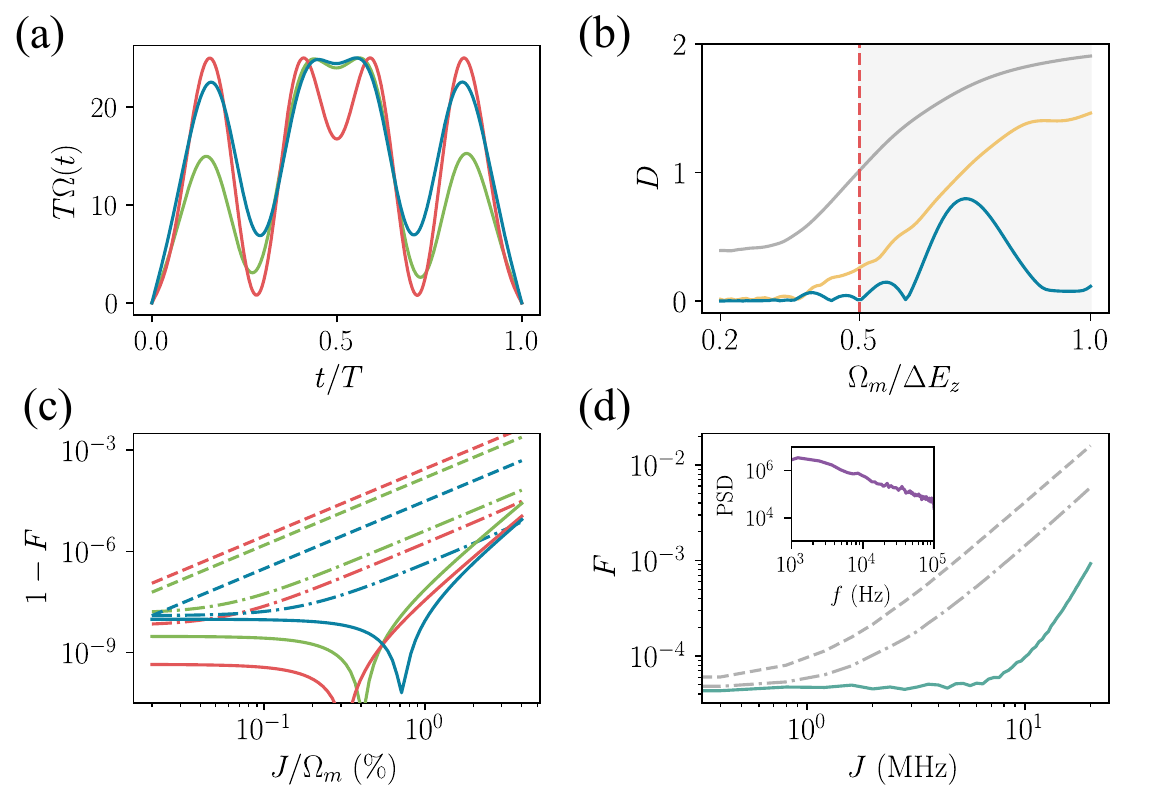}
\caption{ Robust single-qubit gate implementation in a double quantum dot system.
(a) Robust control pulses (RCPs) for the single-qubit gates $\{ X_{\pi}, X_{\pi / 2}, X_{2 \pi} \}$ (blue, red, green). 
(b) Total error distance of the general RCP (blue), a static noise-robust RCP (gold) and a trivial pulse (gray) for $X_{\pi}$ gates as a function of their maximum pulse amplitude $\Omega_m$. The dashed red line marks the working amplitude of the general RCP where $D$ vanishes. As the pulse amplitude increases, crosstalk noise becomes dominant, causing the static noise-robust RCP to lose its effectiveness. Several general RCPs for different working amplitudes are discussed in~\cite{supp}. 
(c) Gate infidelity as a function of the dimensionless always-on coupling
strength $J / \Omega_m$, comparing RCPs (solid) with static noise RCP (dashdot) and trivial pulses (dashed). At large amplitudes, the static noise RCPs degrade and converge with the performance of trivial pulses, see \cite{supp}. 
(d) Gate fidelities for $X_{\pi}$ gates with a gate time of 50 ns, evaluated under $1/f$ frequency noise causing effective $T_2$ about $5$ $\mu\text{s}$. The performance of the general RCP (teal) is compared against a static noise RCP (dash-dotted gray) and a trivial pulse (gray). Inset: noise power spectral density (PSD) of the $1/f$ noise in the frequency range of 1 kHz to 100 kHz.
}
\label{Fig_2Q}
\end{figure}

\paragraph{Single qubit gates.-}

Implementing a single-qubit gate in this two-qubit system corresponds to $d = 1, m = 1$. By combining Eq.~(\ref{Eq_GeneralUe}) and Eq.~(\ref{Eq_2dot_H_Pauli}) and tracing out the spectator qubit, we obtain an effective single-qubit, single-noise Hamiltonian: $\tilde{H}^{1;1}_R = \Omega(t)/2 X + \varepsilon_j Z$ where $\varepsilon_j$ is a noise-dependent parameter given by $\{ J/4, \tan\theta \Omega_2(t)\cos(\Delta E_z t)/2, - \tan\theta \Omega_2(t)\sin(\Delta E_z t)/2 \}$. Our goal is to design an $X$-control that corrects errors arising from these $Z$ noises with small $\varepsilon$. The control pulses are bandwidth-limited and restricted to a set of Fourier components $\Omega_0(a_j, \phi_j; t) = \sin( \pi t/T ) ( a_0 + \sum_{j=1}^n a_j \cos( 2\pi j t/T + \phi_j ) )$. Figure~\ref{Fig_2Q}(a) presents a set of robust control pulses (RCPs) for single-qubit gates $\{ X_{\pi}, X_{\pi/2}, X_{2\pi} \}$, where $X_{\theta}$ denotes a rotation by angle $\theta$ about the x-axis of the Bloch sphere. 
We compare the total error distance of our general $X_{\pi}$ RCP, 
an RCP from prior studies that only addresses static frequency noise and $ZZ$ coupling~\cite{hai2022universal,yi2024robust} and a trivial cosine pulse as a function of their pulse amplitudes in Fig.~\ref{Fig_2Q}(b). The vanishing error distance confirms the first-order noise robustness of the RCPs, while at larger pulse amplitude, the crosstalk effect becomes prominent and the static noise RCP loses its robustness. 
Numerical simulations in Fig.~\ref{Fig_2Q}(c) compare the performance of these RCPs with trivial pulses. The robustness plateau and steeper linear dependence of infidelities on $J$ for the general RCPs confirms their correction of first-order errors, achieving several orders of magnitude improvement in gate fidelity. Figure~\ref{Fig_2Q}(d) further examines gate performance using experiment relevant parameters under time-dependent $1/f$ qubit-frequency fluctuations observed in solid-state platforms~\cite{supp}. 
The general RCP maintains resilience against always-on coupling, while the significant performance gap between it and the other two pulses underscores the importance of suppressing crosstalk noise for robust gate implementation.

\begin{figure}[htb]
	\centering
	\includegraphics[width=\columnwidth]{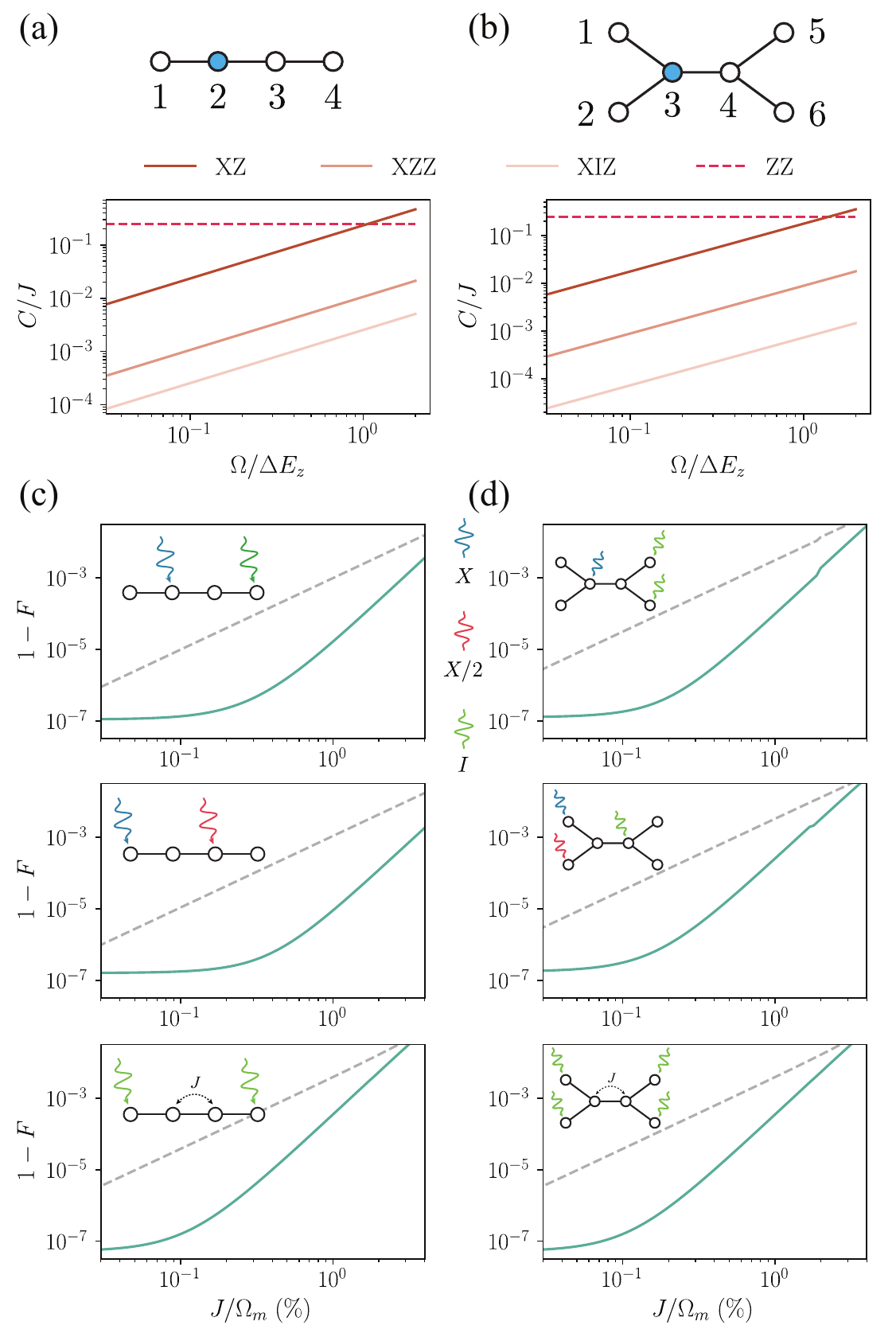}
	\caption{ Analysis of crosstalk and gate fidelity in 1D and 2D multi-qubit systems. (a)(b) Two representative units from the 1D and 2D multi-qubit setups: a four-qubit chain and a six-qubit unit. The relative noise strengths of $ZZ$ interactions (red dashed) are compared with the averaged strengths of two- and three-body crosstalk terms (shades of brown) when driving qubits 2 and 3 with a constant drive amplitude $\Omega$.
    (c) Gate infidelities for the four-qubit system: single-qubit gates $X_{2,\pi}$ and $X_{1,\pi} X_{3,\pi/2}$ (upper panels) and the two-qubit gate ${ZZ}(JT/2)_{23}$ (lower panel) where $JT/2$ is the conditional phase and $T=50$ ns.
    (d) Gate infidelities for the six-qubit system: single-qubit gates $X_{3,\pi}$ and $X_{1,\pi} X_{2, \pi/2}$ (upper panels) and the two-qubit gate $ZZ(JT/2)_{34}$ (lower panel) with $T=50$ ns. In (c)(d), robust and trivial gates are shown in teal and dashed gray, respectively. The pulse colors in the insets correspond to those in Fig.~\ref{Fig_2Q}(a), and $\Omega_m$ is defined as the average maximum absolute amplitude of the pulses used to implement the gates.
	}
	\label{Fig_3_multi_QD}
\end{figure}


\paragraph{Scalable control protocol.-}
Practical quantum computing requires simultaneous control of multiple qubits while maintaining high fidelity~\cite{google2023suppressing}. This becomes particularly challenging in large-scale systems in the presence of non-negligible persistent couplings. To address this, we propose a protocol using robust control for concurrent implementation of universal gates in both one-dimensional chains and two-dimensional honeycomb lattices (Fig.~\ref{Fig_1_architecture}(b)). The unit cell of the 1D chain and 2D honeycomb qubit array are the four and six qubit cells on the control aspect, as shown in Fig.~\ref{Fig_3_multi_QD}(a)(b). Assuming homogeneous coupling strength in the array, we get the effective Hamiltonian from Eq.~(\ref{Eq_GeneralH})
\begin{equation}
\begin{aligned}
\tilde{H}^{m;1} &= \frac{1}{2} \alpha \Omega X_k - \frac{1}{2} \sum_i \tilde{\omega}_i Z_i + \frac{J}{4} \sum_{\langle i,j \rangle} Z_i Z_j \\
&+ \sum_{j}^{\langle j,k \rangle} C^{(2)}_{jk} X_j Z_k
+ \sum_{i,j}^{\langle i,j,k \rangle} C^{(3)}_{ijk} X_i Z_j Z_k
+ ...
\end{aligned}
\end{equation}
where $\alpha$ induces uncertain control inhomogeneity and n-body crosstalk terms $C^{(n)}$ arise for all relevant qubit combinations near the target qubit $k$. This is an extended form of Eq.~\ref{Eq_2dot_H_Pauli} except that here all the coefficients have no analytic form in general. Therefore, we use an exact block diagonalization~\cite{guan2025photon} to solve for the relative amplitudes of crosstalk $C^{(n)}$ in the unit cell. As shown in Fig.~\ref{Fig_3_multi_QD}(a)(b), nearest-neighbor ${XZ}$ crosstalk dominates and approaches the strength of the always-on coupling under strong driving. The higher-order terms are several orders weaker and are negligible in the rotating frame. In fact, our simulation shows that including these higher-order terms in the calculation produces no appreciable difference in gate fidelity. 

As demonstrated in the two-qubit scenario, the implementation of robust single-qubit gates facilitates the correction of errors affecting the immediate neighboring qubits and inhibits the progression of errors to subsequently distant neighbors. As a logical progression, a protocol has been developed to execute robust single-qubit gates through the application of robust control pulses (RCPs) on alternating qubits, thereby effectively correcting ${ZZ}$ interactions and crosstalk. Fig.~\ref{Fig_3_multi_QD}(c)(d) shows the infidelities of parallel single-qubit gates, including $X_{\pi}$ and $X_{\pi/2}$ gates, in the unit cell of 1D chain and 2D honeycomb qubit arrays. Universal single-qubit control can be implemented using only robust $X_{\pi/2}$ and $X_{-\pi/2}$ gates with one robust pulse plus virtual-$Z$ rotations, following Euler decomposition $U(\alpha,\beta,\lambda) = Z_\beta X_{\pi/2} Z_\alpha X_{-\pi/2} Z_\lambda$. As demonstrated in the last column of
Fig.~\ref{Fig_3_multi_QD}(c)(d), two-qubit gates emerge naturally from the always-on coupling, with identity gates $X_{2\pi}$ using RCPs on the adjacent qubits to sufficiently separate the entangling operation.

\begin{figure}[tb]
	\centering
	\includegraphics[width=\columnwidth]{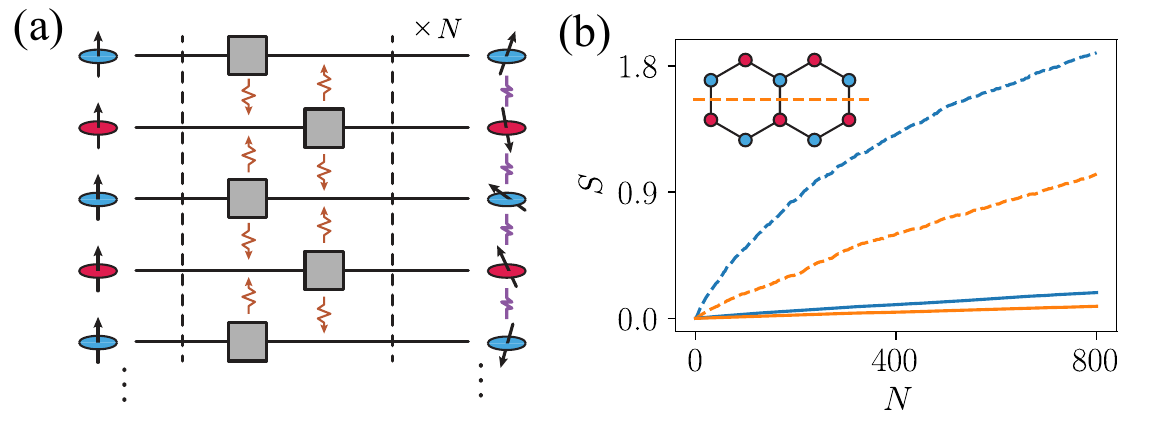}
	\caption{ Numerical investigation of entanglement dynamics in a 2D qubit system with always-on coupling. (a) Schematic of a multi-qubit quantum circuit with $N$ layers of randomly selected single-qubit gates (gray boxes). Wavy lines with arrows indicate error propagation through always-on coupling, leading to entanglement growth despite the circuit containing only single-qubit gates and potentially resulting in chaotic dynamics. (b) Comparison of entanglement entropy evolution using robust gates (solid) versus trivial gates (dashed). The curves represent bipartite entanglement entropy for an even-odd partition (blue) and an upper-lower partition (orange), corresponding to the color-coded regions in the inset. The data is averaged over 100 random realizations with circuit depths up to 800 layers. The initial state is all spins down, other product initial states exhibit similar behavior. }
	\label{Fig_Entropy}
\end{figure}

\paragraph{Preventing chaos via robust gates.}
Large-scale correlated errors can pose a major challenge to fault-tolerant quantum computing by introducing chaotic dynamics and entanglement growth that leads to additional decoherence in qubits. Always-on coupling in transmon-based processors and spin qubit systems has been shown to introduce chaotic fluctuations that destabilize qubit states and increase computational errors~\cite{berke2022transmon, borner2024classical, banuls2011strong, gubin2012quantum, kim2014testing}. These effects intensify with system size, as larger qubit arrays and higher-dimensional architectures accelerate the onset of chaos due to enhanced connectivity and delocalization. Figure~\ref{Fig_Entropy}(a) illustrates this error propagation, where deep quantum circuits lead to chaotic entanglement and obscure useful quantum information. While mitigation strategies such as engineered disorder and tunable coupling offer partial solutions, they fail to fully suppress delocalization and chaos, particularly in large, complex systems. Floquet control has been proposed to address these issues~\cite{mukherjee2024arresting}, but conventional gate operations in multi-qubit systems with always-on coupling often introduce additional errors~\cite{yi2024robust} and further reduce performance. This highlights the urgent need for scalable, robust control mechanisms to correct spatially correlated errors and prevent chaotic dynamics.

We demonstrate that robust pulse sequences effectively suppress chaotic behavior in multi-qubit systems with always-on couplings. In ideal quantum circuits, single-qubit gates cannot generate entanglement between qubits. However, in systems with always-on couplings, even basic single-qubit operations can induce unwanted entanglement and eventually lead to chaotic dynamics, fundamentally limiting circuit fidelity and coherence time. As shown in Fig.~\ref{Fig_Entropy}(b), we track the progression of entanglement entropy as a function of circuit depth in quantum circuits composed of repeated layers of either conventional single-qubit gates (dashed lines) or robust gates (solid lines), randomly selected from the set $\{X_{\pi}, X_{\pi / 2}, X_{2 \pi} \}$. We evaluate two types of bipartite entanglement entropy, $S(\rho_A)=-\operatorname{Tr}(\rho_A\log\rho_A)$, where $\rho_A$ is the reduced density matrix of subsystem $A$ and $A$ represents either the odd or upper half-system. Simulations of a 2D $10$-qubit system show that conventional gates lead to a rapid increase in entanglement entropy, reflecting the accumulation of correlated errors. In contrast, circuits using robust gates significantly slow entanglement growth and prevent error propagation. 
These results confirm that robust pulses effectively suppress chaotic dynamics, providing a crucial tool for scaling multi-qubit systems with always-on couplings.

\paragraph{Conclusion.-}We have demonstrated a robust control scheme for semiconductor spin qubit systems with always-on couplings that addresses a key scaling challenge: correlated errors and crosstalk-induced chaos. Our numerical simulations show that robust control pulses can implement high-fidelity quantum gates robust against both static coupling noise and time-dependent crosstalk effects. Importantly, these robust gates mitigate unwanted entanglement growth, preserving quantum coherence in large multi-qubit systems even through deep quantum circuits.

Our approach is scalable to various qubit architectures, including one-dimensional chains and two-dimensional honeycomb lattices, without requiring high on/off ratio tunable couplers. This significantly relaxes hardware constraints for silicon-based quantum processors and hence has substantial implications. For some systems, directly achieving high controllability of $J$ is physically challenging with current technologies, as exemplified by the 1P-1P system in phosphorus donors~\cite{Wang2016-rs}. Additionally, several silicon systems encounter difficulties in attaining a high dynamic range of $J$ within practical gate voltage parameter sets~\cite{xue2022quantum, noiri2022fast, Mills2022, Huang2024-fz}. A more accessible tuning range for $J$ implies reduced demands on the precision of $J$ gate fabrication and lithographic line widths, potentially resulting in higher yields using moderate fabrication techniques, thereby making chip production more economical~\cite{neyens2024probing, steinacker2024300mm}. Furthermore, our proposal even allows for a fixed $J$ coupling, reducing the number of required $J$ electrodes and significantly reducing fan-out overhead. In architectures such as crossbar~\cite{Li2018} or shared control designs~\cite{Hill2015, Borsoi2024}, reduced tunability requirements lead to increased qubit yield and feasibility, further facilitating the scaling of the system with state-of-the-art fabrication technologies.

By providing a promising path to large-scale quantum information processing that works within the constraints of current fabrication capabilities, our control strategy represents a viable approach for scaling semiconductor quantum processors. The techniques developed here may find applications in other densely coupled qubit platforms where always-on interactions present similar challenges.

\paragraph{Acknowledgements.} We thank inspiring suggestions from Liang Jiang in the early stages of this project and fruitful discussions with Junkai Zeng in the scalable protocol. This work was supported by the National Natural Science Foundation of China (Grants No. 92165210, 62174076), the Key-Area Research and Development Program of Guang-Dong Province (Grant No. 2018B030326001), the Innovation Program for Quantum Science and Technology (No. 2021ZD0302300), and the Science, Technology and Innovation Commission of Shenzhen Municipality (JCYJ20170412152620376, KYTDPT20181011104202253), and the Shenzhen Science and Technology Program (KQTD20200820113010023). 

\bibliographystyle{apsrev4-1}
\bibliography{Reference}

\clearpage
\thispagestyle{empty}  
\newpage 




\begin{titlepage}
\begin{center}
\large\textbf{Supplementary for "Scalable Robust Quantum Control for Semiconductor Spin Qubits with Always-on Couplings"}

\vspace{1cm}
\normalsize
Yong-Ju Hai$^{1}$, Shihang Zhang$^{2}$, Haoyu Guan$^{2}$, Peihao Huang$^{1}$, Yu He$^{1,2}$, Xiu-Hao Deng$^{1,2,*}$

\vspace{0.5cm}
\small
$^{1}$Shenzhen International Quantum Academy (SIQA), Futian District, Shenzhen, P. R. China\\
$^{2}$Shenzhen Institute for Quantum Science and Engineering (SIQSE), Southern University of Science and Technology, Shenzhen, P. R. China\\
$^{*}$Email: dengxh@sustech.edu.cn
\end{center}
\end{titlepage}

\section{Quantum dot qubit model}

In this section, we detail the theoretical model used to describe the semiconductor spin qubits, which form the fundamental building block of our scalable robust quantum control scheme.

\subsection{Two-qubit Model}

We first consider a pair of quantum dot spin qubits operating in the $(1,1)$ charge configuration, where $(n_1, n_2)$ denotes the number of electrons in each dot. The system is governed by an extended Heisenberg model \cite{burkard2023semiconductor}: \begin{equation}
	H=\mathbf{B}_{1} \cdot \mathbf{S}_{1}+\mathbf{B}_{2} \cdot \mathbf{S}_{2}+J\left(\mathbf{S}_{1} \cdot \mathbf{S}_{2}-1 / 4\right).
	\label{Eq_Heisenberg_Hamiltonian}
\end{equation}
where $S_j = (X_j, Y_j, Z_j)/2$ are spin operators (in terms of Pauli matrices \{$X,Y,Z$\}, and $B_j = (B_{x,j}, B_{y,j}, B_{z,j} )$ are the magnetic fields at each qubit site, and $J$ denotes the exchange coupling between spins.

The $z$-components of the magnetic fields determine distinct electron spin resonance (ESR) frequencies, allowing for selective qubit control. Transverse control fields ($B_{x,j}$, $B_{y,j}$) are generated by gate voltages, while $J$ is modulated by the inter-dot barrier gate.

Defining the average Zeeman energy and Zeeman energy difference as $E_z = \frac{1}{2} (B_{z,1} + B_{z,2})$ and $\Delta E_z = B_{z,2} - B_{z,1}$, we write the Hamiltonian without transverse drive in the spin basis $\{\left| \uparrow \uparrow \right\rangle, \left| \uparrow \downarrow \right\rangle, \left| \downarrow \uparrow \right\rangle, \left| \downarrow \downarrow \right\rangle \}$ as 
\begin{equation}
	H_0 = \frac{1}{2} \begin{pmatrix}
		2E_z & 0 & 0 & 0 \\
		0 & -\Delta E_z - J & J & 0 \\
		0 & J & \Delta E_z - J & 0 \\
		0 & 0 & 0 & -2E_z
	\end{pmatrix}.
\end{equation}
Diagonalizing this Hamiltonian yields
\begin{equation}
	\tilde{H}_0 = \frac{1}{2} \begin{pmatrix}
		2E_z & 0 & 0 & 0 \\
		0 & - \Delta \tilde{E}_z - J & 0 & 0 \\
		0 & 0 & \Delta \tilde{E}_z - J & 0 \\
		0 & 0 & 0 & -2E_z
	\end{pmatrix}.
\end{equation}
in the eigenbasis $\{\left|\uparrow \uparrow \right\rangle, |\tilde{\uparrow \downarrow}\rangle, |\tilde{\downarrow \uparrow}\rangle, \left|\downarrow \downarrow \right\rangle\}$, where $|\tilde{\uparrow \downarrow}\rangle=\cos\theta|\uparrow \downarrow\rangle+\sin\theta| \downarrow\uparrow\rangle$, $ |\tilde{\downarrow \uparrow}\rangle=\sin\theta|\uparrow \downarrow\rangle+\cos\theta| \downarrow\uparrow\rangle$ and $\tan \theta = \frac{J}{\Delta E_z + \sqrt{J^2 + \Delta E_z^2}}$, $\Delta \tilde{E}_{z}=\sqrt{J^{2}+\Delta E_{z}^{2}}$.

Adding a $x$-magnetic field on spin 2, the Hamiltonian in the eigenbasis becomes
\begin{widetext}
	\begin{equation}
		\tilde{H}=\frac{1}{2}\left(\begin{array}{cccc}
			2 E_{z} & \cos \theta B_{x,2} & \sin \theta B_{x,2} & 0 \\
			\cos \theta B_{x,2}^* & -\Delta \tilde{E}_{z}-J & 0 & - \sin \theta B_{x,2} \\
			\sin \theta B_{x,2}^* & 0 & \Delta \tilde{E}_{z}-J & \cos \theta B_{x,2} \\
			0 & - \sin \theta B_{x,2}^* & \cos \theta B_{x,2}^* & -2 E_{z}
		\end{array}\right).
		\label{Eq_2dot_H1_full}
	\end{equation}
\end{widetext}
Transforming into the rotating frame using $\tilde{H}_R = R^{\dagger} \tilde{H} R + i R^{\dagger} \partial_t R$, with $R = e^{i \frac{1}{2}(E_z - \frac{1}{2} \Delta \tilde{E}_z) t ZI + i \frac{1}{2}(E_z + \frac{1}{2} \Delta \tilde{E}_z) t IZ}$. We then choose the drive field that satisfy $\cos \theta B_{x,2} = \Omega_2 e^{i \omega_2 t}$ to compensate the cosine factor, with the effective control amplitude $\Omega_2$ and frequency $\omega_2 = E_z + \Delta \tilde{E}_z/2$ to get the rotating frame Hamiltonian
\begin{widetext}
	\begin{equation}
		\tilde{H}_R = \frac{1}{2}\left(\begin{array}{cccc}
			J/2 & \Omega_{2} & e^{i\tilde{E}_z t} \tan \theta \Omega_{2} & 0 \\
			\Omega_{2} & -J/2 & 0 & - e^{i\tilde{E}_z t} \tan \theta \Omega_{2} \\
			e^{-i\tilde{E}_z t} \tan \theta \Omega_{2} & 0 & -J/2 &  \Omega_{2} \\
			0 & - e^{-i\tilde{E}_z t} \tan \theta \Omega_{2} &  \Omega_{2} & J/2
		\end{array}\right) 
		= \begin{pmatrix}
		\tilde{H}_1 + \tilde{V}_{11} & \tilde{V}_{12} \\
		\tilde{V}_{21} & \tilde{H}_2 + \tilde{V}_{22}
		\end{pmatrix}.
		\label{Eq_2dot_H_rot}
	\end{equation}
\end{widetext}
This Hamiltonian has a block form. The target qubit has two raw Hamiltonians
\begin{equation}
\tilde{H}_{1,2} = \frac{1}{2} \begin{pmatrix}
	\pm J/2 & 0 \\
	0 & \mp J/2
\end{pmatrix}
\end{equation}
and the control terms
\begin{equation}
\tilde{V}_{11} = \tilde{V}_{22} = \frac{1}{2} \begin{pmatrix}
	0 & \Omega_2 \\
	\Omega_2 & 0
\end{pmatrix}
\end{equation}
according to the state of the first qubit coupled to it. The off-diagonal block that connects the two subspaces
\begin{equation}
\tilde{V}_{12,21} = \frac{1}{2} \begin{pmatrix}
	e^{\pm i\tilde{E}_z t} \tan \theta \Omega_2 & 0 \\
	0 & -e^{\pm i\tilde{E}_z t} \tan \theta \Omega_2
\end{pmatrix}
\end{equation}
represents the quantum crosstalk arising from the coupling, which leads to conditional drives on qubit 1. The Hamiltonian can be written in the following compact form
\begin{equation}
	\begin{aligned}
		\tilde{H}_R &= \frac{\Omega_2(t)}{2} IX + \frac{J}{4} ZZ
		+ \frac{1}{2} \tan\theta \Omega_2(t) \\ 
		& \times
		\Big[\cos(\Delta \tilde{E}_z t) XZ - \sin(\Delta \tilde{E}_z t) YZ \Big],
	\end{aligned}
	\label{Eq_2dot_H_Pauli}
\end{equation}
where the always-on coupling leads to a static $ZZ$ coupling and two quantum crosstalk terms \{$XZ$, $YZ$\}.

\subsection{Multi-qubit Model}

In multi-qubit systems, such as one-dimensional chain and a two-dimensional honeycomb lattice discussed in the main text, we have the Hamiltonian extended from Eq.\ref{Eq_Heisenberg_Hamiltonian}
\begin{equation}
	H_0 =  - \sum_i \frac{\omega_i}{2} Z_i +  \sum_{\langle i,j \rangle} \frac{J_{ij}}{4} (X_i X_j + Y_i Y_j + Z_i Z_j),
\end{equation}
where $\langle i, j \rangle$ denotes neighboring qubit pairs and we take homogeneous coupling strength $J_{ij} = J$ for simplicity.

We perform numerical diagonalization of the few qubit building blocks of multi-qubit systems and observe the nearest neighbor crosstalk is the dominant noise source. So we consider a simplified two-qubit model with neighboring crosstalk. Given qubit eigen-frequencies $\omega_1$ and $\omega_2$, which can be experimentally calibrated, we apply a transverse drive to qubit 2 at frequency $\omega_2$, along with a transverse crosstalk term. The Hamiltonian is
\begin{equation}
	\begin{aligned}
		H = &-\frac{\omega_1}{2} Z_1 - \frac{\omega_2}{2} Z_2 
		+ \frac{\Omega_{2}}{2} ( e^{-i\omega_2 t}\sigma_2^+ + e^{i\omega_2 t}\sigma_2^- ) 
		\\ &+ \beta \Omega_{2}( e^{-i\omega_2 t}\sigma_1^+ + e^{i\omega_2 t}\sigma_1^- )Z_2 + \frac{J}{4} Z_1 Z_2,
	\end{aligned}
\end{equation}
where $\beta$ is a constant factor. Transforming into the rotating frame using $R = e^{-i(\frac{1}{2} \omega_1 Z_1 + \frac{1}{2} \omega_2 Z_2 )}$, we obtain
\begin{equation}
	\begin{aligned}
		H_R & = \frac{\Omega_{2}}{2} X_2 + \frac{J}{4} Z_1 Z_2 \\
		& + \beta \Omega_{2} (\cos\Delta_{12}t X_1 + \sin\Delta_{12}t Y_1)Z_2
	\end{aligned}
\end{equation}
where $\Delta_{12} = \omega_2 - \omega_1$. This Hamiltonian follows the same structure as Eq.\ref{Eq_2dot_H_Pauli}. In multi-qubit systems with multiple crosstalk terms, this model extends naturally, and the robust control pulses we obtained for the original two-qubit model remain applicable.

\section{Robust control pulses}
\subsection{Pulse construction protocol}

\begin{figure*}[hbt!]
	\centering
	\includegraphics[width=1\textwidth]{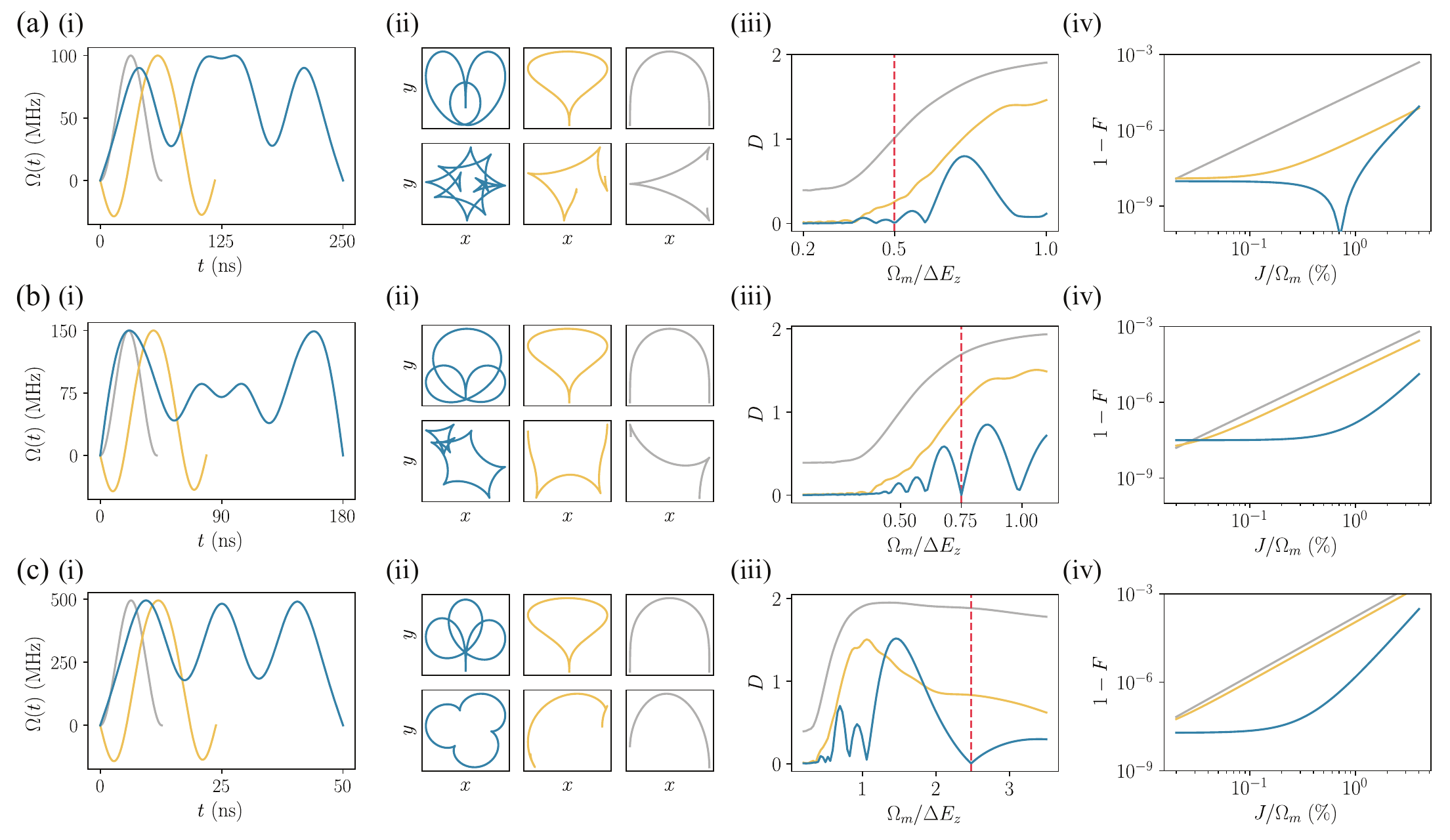}
	\caption{ (a)-(c) Properties of robust control pulses $X_{\pi}$ at different working amplitudes/gate times. 
		(i) Illustration of RCPs (blue), static noise RCPs (gold) and trivial cosine pulses (gray). All pulses are normalized to the same maximum amplitude. 
		(ii) The error curves of the three pulses for $ZZ$ coupling (upper panels) and $ZY$ crosstalk (lower panels). The general RCPs are robust against both noises and have two closed error curves. The static noise RCPs are only robust against $ZZ$ coupling and have one closed error curve. The trivial pulses have two open curves and are not robust. 
		(iii) The total error distance of the three noises ($ZZ$ coupling and $ZX$, $ZY$ crosstalks) as a function of different amplitude obtained by linear scaling $\Omega \to \Omega/a$, $t \to a t$. The dashed red line marks the working amplitudes of the general RCPs. 
		(iv) Gate infidelities as a function of dimensionless coupling strength. As amplitude increases, crosstalk noise becomes more prominent and the static noise RCPs become less robust and eventually converge with the performance of trivial pulses, which agree with the error distance in (iii).
	}
	\label{Fig_compare_pi_pulses}
\end{figure*}

To establish a protocol to construct robust pulse, we first analyze the error dynamics generated by the noises. Consider system Hamiltonian Eq.\ref{Eq_2dot_H_Pauli}, the total evolution operator can be decomposed as $U = U_0 U_{\text{e}}$, where $U_0(t) = \mathcal{T} \exp\{-i \int_0^t d\tau H_{0}(\tau)\}$ is the ideal evolution operator and $U_{\text{e}}(t) = \mathcal{T} \exp\{-i \int_0^t d\tau \tilde{V}_{\text{I}}(\tau)\}$, where $\tilde{V}_{\text{I}}= U_0^{\dagger} \tilde{V} U_0$ represents the noise Hamiltonian in the interaction picture with
\begin{equation}
	\begin{aligned}
		H_0 & = \frac{1}{2} \Omega_2(t) IX \\
		V & = \frac{J}{4} ZZ
		+ \frac{1}{2} \tan\theta \Omega_2(t) \\ 
		& \times
		\Big[\cos(\Delta \tilde{E}_z t) XZ - \sin(\Delta \tilde{E}_z t) YZ \Big].
	\end{aligned}
\end{equation}
Up to first perturbative order, different noises don't interfere with each other and we consider the impact of them separately, which leads to a further separation of error unitary $U_e = \prod_j U_{e,j}$, where $i$ labels different noise sources and $U_{e,j}(t) = e^{-i \int_0^t d\tau U_0^{\dagger}(\tau) \tilde{V}_j(\tau) U_0(\tau) }$, where $\tilde{V}_j$ represents different noise terms. In this case, each error evolution is restricted to the subspace generated by the control term and the corresponding noise term $\{IX,KZ\}$ with $K = Z$ or $K = X, Y$ corresponding to the $ZZ$ noise and crosstalks, respectively.

As a consequence, we can trace out qubit 1 to obtain effective single-qubit control and noise Hamiltonians for qubit qubit 2 
\begin{equation}
	\begin{aligned}
		H_0 & = \frac{1}{2} \Omega(t) X \\
		V_1 & = \frac{J}{4} Z = \epsilon_1 Z \\
		V_2 & = \frac{1}{2} \tan\theta \Omega_2(t) \cos(\Delta \tilde{E}_z t) Z 
		= \epsilon_2 v_2(t) Z \\
		V_3 & = - \frac{1}{2} \tan\theta \Omega_2(t) \sin(\Delta \tilde{E}_z t) Z 
		= \epsilon_3 v_3(t) Z
		\label{Eq_1Q_H_eff}
	\end{aligned}
\end{equation}
where $\epsilon_j$ are the constant factors (we ignore the $J$-dependence on $\theta$ at first order), $v_j(t)$ are time-dependent profiles for each noise. 
The error dynamics generated by $V_1$ to $V_3$ happen in different subspaces. The noise susceptibility is quantified by error distance $|\mathbf{r}_j|$ defined using the final error unitary at gate time $T$, $U_{e,j}(T) = e^{-i \int_0^T d\tau U_0^{\dagger}(\tau) \tilde{V}_j(\tau) U_0(\tau) } = e^{-i \epsilon_j \mathbf{r}_j \cdot \sigma}$. Here $U_0$ is the single-qubit ideal unitary generated by the effective control Hamiltonian $H_0$ in Eq.\ref{Eq_1Q_H_eff}, $\sigma$ is the Pauli matrix vector. Specifically, we have
\begin{equation}
	\begin{aligned}
		\mathbf{r}_1(t) \cdot \sigma & =  \int_0^t d\tau U_0^{\dagger} Z U_0\\
		\mathbf{r}_2(t) \cdot \sigma & = \int_0^t v_1(\tau) U_0^{\dagger} Z U_0 \\
		\mathbf{r}_3(t) \cdot \sigma & = - \int_0^t v_2(\tau) U_0^{\dagger} Z U_0 .
		\label{Eq_error_curves}
	\end{aligned}
\end{equation}
We then define the total error distance $D = \sum_j \| \mathbf{r}_j(T) \|$ as a metric of control robustness. 

\begin{figure*}[tb]
	\centering
	\includegraphics[width=1\textwidth]{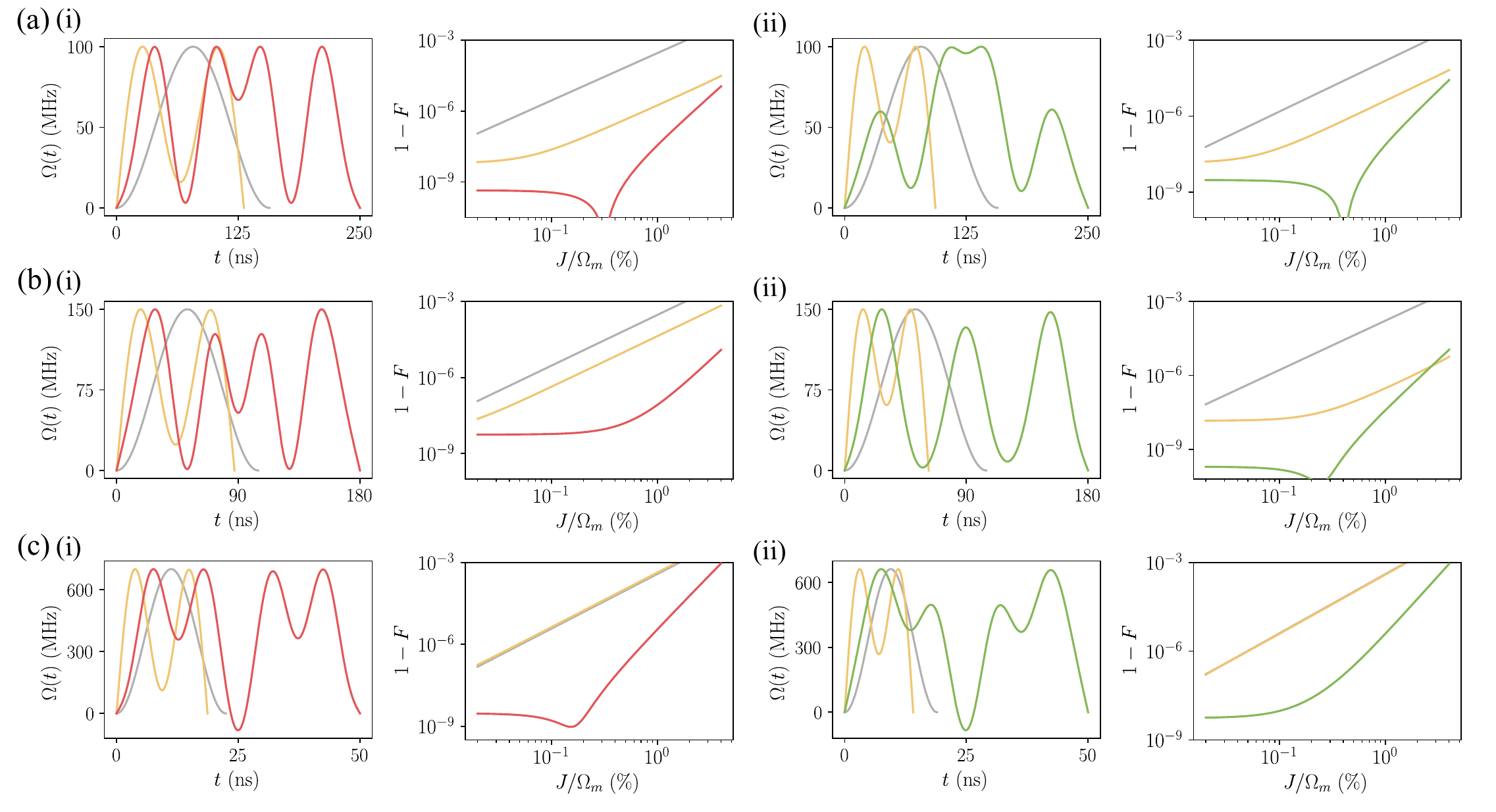}
	\caption{ (a)-(c) Robust control pulses $X_{\pi/2}$ and $X_{2\pi}$ at different working amplitudes/gate times. 
	(i)(ii) Left panel: Illustration of $X_{\pi/2}$ and $X_{2\pi}$ RCPs (red, green), static noise RCPs (gold) and trivial cosine pulses (gray). All pulses are normalized to the same maximum amplitude. 
	(i)(ii) Right panel: Gate infidelities of the three pulses. As amplitude increases, the static noise RCPs become less robust and converge with the performance of trivial pulses, agreeing with the result in Fig.~\ref{Fig_compare_pi_pulses}.
	}
	\label{Fig_compare_pulses}
\end{figure*}

We use a numerical pulse construction protocol based on autodifferentiation similar to that used in \cite{hai2022universal}. The pulses are parameterized as a modified Fourier series
\begin{equation}
	\begin{aligned} 
		\Omega(a_j,\phi_j;t) &= \sin( \frac{\pi t}{T} )(a_0 +
		\sum_{j=1}^n a_j \cos( \frac{2\pi j}{T} t + \phi_j) )
	\end{aligned}  \label{Eq_Pansatz}
\end{equation}
with fixed total gate time $T$ and the number of Fourier components $n$. $\{ a_j,\phi_j, b_j,\psi_j\}$'s are parameters to be optimized. The Fourier form features smoothness and limited bandwidth to be experimentally friendly, and the sine coefficient ensures the pulses start and end at zero amplitude. The protocol is described as follows:

(1) Initialize the input parameters.

(2) Apply a linear pulse amplitude constraint by multiplying the input pulse by a scale factor $u/\Omega_m$, to scale the pulse amplitude within $[-u,u]$, where $\Omega_m$ is the maximum amplitude of the input pulse.

(3) Use the modified pulses to compute the noiseless dynamics with Hamiltonian $H_0 = \Omega(t)/2 X$ to obtain the evolution operator $U_0(t)$ and then calculate ideal gate fidelity $F$ and total error distance $D$. 

(4) Compute the cost function
\begin{equation}
	C=(1-F) + D ,
	\label{Eq_CostF}
\end{equation}%

(5) Make a gradient update of the pulse parameters in order to minimize $C$.

(6) Go back to step (2) with the updated pulse parameters as input if the cost function is larger than a criterion $\eta$.

(7) If $C < \eta$, break the optimization cycle and obtain the optimal robust pulse.

We obtained three sets of robust control pulses at gate times $\{ 50, 180, 250\}$ ns and maximum amplitudes at about $\{ 100, 150, 600\}$ MHz ($\Delta E_z = 200$ MHz). The pulse parameters are listed in the following table~\ref{Table_Pulse}.

\begin{table*}[htb]
	\caption{Parameters for the RCPs presented in this work. The pulses have the analytical form Eq.~(\ref{Eq_Pansatz}) and the amplitude unit is GHz. The key parameters are the relative amplitude $\Omega_m/\Delta E_z$ with $\Delta E_z = 200$ MHz, gate time $T$ (ns), amplitude parameter $a_j$'s and phase parameter $\phi_j$'s. }
	\begin{tabular}{lllll}
\hline\hline
RCPs & $\Omega_m/\Delta E_z$ & $T$ & $a$ & $\phi$  \\
\hline
$X_{\pi}$ & 100 & 250 & [0.1225, 0.0672, 0.0394, -0.0297, -0.0228,
0.0040] & [0.0022, -0.0138, 0.0028, 0.0114, -0.0595] \\
\hline
$X_{\pi/2}$ & 0.5 & 250 & [0.1067, 0.0547, 0.0261, -0.0470, -0.0538,
0.0044]  & [0.0016, 0.0069, -0.0067, -0.0050, 0.0078] \\
\hline
$X_{2\pi}$ & 0.5 & 250 & [0.0906, 0.0292, 0.0429, -0.0188, -0.0255, 0.0017] & [-0.0082, 0.0114, 0.0056, 0.0448, -0.2691] \\
\hline
$X_{\pi}$ & 0.75 & 180 & [0.2374, 0.2683, 0.1459, 0.0335, 0.0030, 0.0144] & [-0.0055, -0.0021, -0.0006, -0.2457, -0.0157] \\
\hline
$X_{\pi/2}$ & 0.75 & 180 & [0.1735, 0.1438, 0.0625, -0.0427, -0.0606, 0.0207] & [0.0013, 0.0049, -0.0139, -0.0093, 0.0062] \\
\hline
$X_{2\pi}$ & 0.75 & 180 & [0.1522, 0.1288, 0.0434, -0.0866, -0.0375, -0.0174] & [0.0093, -0.0431, 0.0567, -0.0104, -0.0313] \\
\hline
$X_{\pi}$ & 2.5 & 50 & [0.6191, 0.3799, 0.0626, -0.1812, -0.0006, -0.0001] & [-0.0027, -0.0669, -0.0056, 0.0041, 0.0111] \\
\hline
$X_{\pi/2}$ & 3.5 & 50 & [0.7961, 0.5159, -0.1174, -0.0838, -0.4011, -0.0727] & [0.0013, -0.0085, -0.0026, 0.0043, -0.0586] \\
\hline
$X_{2\pi}$ & 3.3 & 50 & [0.8686, 0.8161, 0.1008, 0.0318, -0.2056, -0.0007] & [-0.0049, -0.0994, -0.1188, 0.0682, 0.1152] \\
\hline\hline
\end{tabular}
\label{Table_Pulse}
\end{table*}

\subsection{Robust control pulses in different parameter regime}

In this section, we compare the performance of different general robust control pulses (RCPs) obtained in this work, the static noise RCPs only robust against static frequency noise and $ZZ$ coupling studied in~\cite{hai2022universal,yi2024robust} and the trivial cosine pulses. 

In Fig.~\ref{Fig_compare_pi_pulses}, we show $X_{\pi}$ RCPs at gate times $\{ 50, 180, 250\}$ ns and maximum amplitudes at about $\{ 100, 150, 500\}$ MHz ($\Delta E_z = 200$ MHz) and calculate the error curves $\mathbf{r}_1(t)$ and $\mathbf{r}_3(t)$ from Eq.~\ref{Eq_error_curves}, which corresponds to the error evolution of $ZZ$ coupling and $ZY$ crosstalk, respectively. The closeness of both error curves for RCPs confirms their robustness against both noises. In contrast, for static noise RCPs and simple cosine pulses, the total error distance increases with pulse amplitude due to the growing impact of crosstalk noise. As a result, the performance of static noise RCPs degrades and eventually approaches that of trivial pulses. Meanwhile, the general RCPs maintain their robustness, highlighting the necessity of simultaneously suppressing both noise sources at high pulse amplitudes. We also compare the gate fidelities of the three pulse types for $X_{\pi/2}$ and $X_{2\pi}$ gates in Fig.~\ref{Fig_compare_pulses}, and observe consistent conclusions.

\subsection{Simulation of Low-frequency Noise}

\begin{figure}[htb!]
	\centering
	\includegraphics[width=\columnwidth]{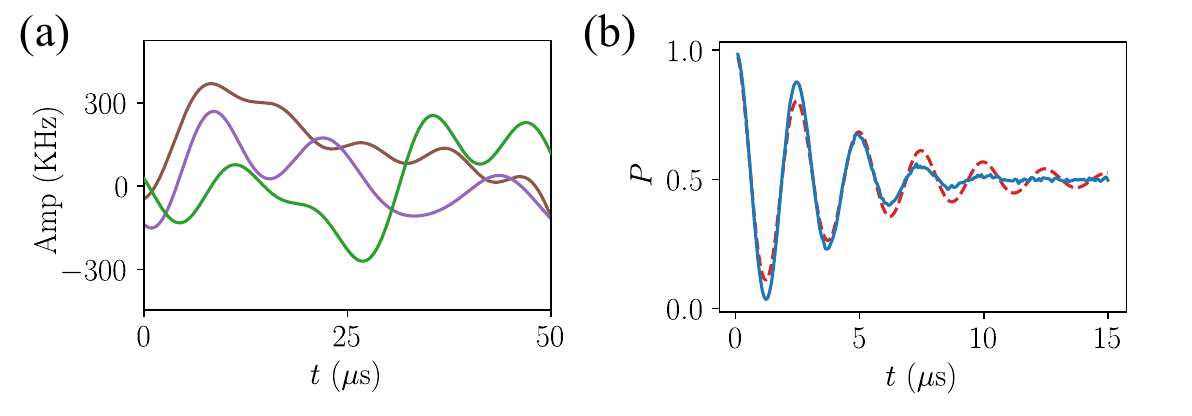}
	\caption{ (a) Typical single-shot time series samples of $1/f$ qubit frequency noise. 
	(b) Simulated decoherence time $T_2$ measurement via Ramsey experiment. Solid: simulation in an average of $5000$ noise instances. Dashed: Fitted curve with $T_2 = 5$ us. 
	}
	\label{Fig_A1_LF_noise}
\end{figure}

In reality, despite the noise in coupling strength and control crosstalk considered in the main text, another dominant noise source that has been widely observed in solid-state platforms is the time-dependent low-frequency fluctuations in qubit frequency, often characterized by a power-law spectrum, such as 
$1/f$. To more faithfully predict the performance of our RCPs under realistic conditions, we study the behavior of RCP gates in a double quantum dot system where the qubits are subject to frequency fluctuations with a $1/f$ spectrum.

We generate the time series of $1/f$ using a combination of sine waves~\cite{timmer1995generating}.
\begin{equation}
   \delta(t) = \gamma \sum_i\sqrt{\frac1{f_i}}\sin(2\pi f_it+\phi_i) 
\end{equation}
where $\gamma$ is the amplitude parameter, the frequencies $f_i$ are chosen within a desired range $f_i \in [f_\text{min}, f_\text{max}]$ and $\phi_i$ are random phases in $[0, 2\pi]$. We consider the frequency ranges of $1-100$ kHz and the noise amplitude $\gamma = 10^6$, chosen such that the frequency fluctuation can reach a few hundred kHz, as shown in Fig.~2(d) and Fig.~\ref{Fig_A1_LF_noise}(a). This choice of parameters is in line with the typical experimentally measured values of long-term fluctuations in spin qubits' frequencies \cite{watson2018programmable, connors2022chargenoise,xue2022quantum}, leading to an $5$ us effective decoherence time (Fig.~\ref{Fig_A1_LF_noise}(b)).

We consider the system Hamiltonian from Eq.~(\ref{Eq_2dot_H_Pauli}) with additional frequency noise applied to both qubit in the double quantum dot and compare the performance of the general RCP demonstrated in this work, the frequency noise RCP from~\cite{hai2022universal}, and a conventional cosine pulse to implement $X_{\pi}$ gates on Q2, as shown in Fig.~2(d). Compared to the results shown in Fig.~2(c), although the highest gate fidelity is reduced due to low-frequency noise and qubit decoherence, the robustness of the RCP against variations in coupling strength remains intact. While the frequency noise RCP exhibits robustness against frequency noise, it is not resilient to crosstalk noise at finite coupling strengths. This result highlights the importance of correcting crosstalk noise using more sophisticated RCPs.

\section{Experimental Relevance of Robust Control Protocol}

\begin{table*}[htb]
	\caption{Parameters for state-of-the-art semiconductor spin qubit devices based on gate-defined quantum dots in Si and Ge, and on P-donors in Si. The effective driving strength is calculated by taking reciprocal of the single-qubit gate time.}
\begin{tabular}{|c|c|c|c|c|c|c|}
\hline
Reference & Xue~\cite{xue2022quantum} & Noiri~\cite{noiri2022fast} & Philips~\cite{philips2022universal} & Hendrickx~\cite{Hendrickx2020N} & Wang~\cite{WangCA2024} & Stemp~\cite{Stemp2024} \\
\hline
Platform & Si/SiGe & Si/SiGe & Si/SiGe & Ge/SiGe & Ge/SiGe & Si:P \\
\hline
Single-qubit gate time & 0.2 $\mu$s & 0.2 $\mu$s & 0.2 $\mu$s & 0.04 $\mu$s & 0.38/0.14 $\mu$s & 1 $\mu$s \\
\hline
Driving stength $\Omega$ & 5 MHz & 5 MHz & 5 MHz & 25 MHz & 2.6/7.1 MHz & 1 MHz \\
\hline
$J_{\mathrm{on}}$ & $\sim$10 MHz & $\sim$20 MHz & $\sim$10 MHz & 39 MHz & 40 MHz & 12 MHz \\
\hline
$J_{\mathrm{off}}$ & 20-100 kHz & / & 15-39 kHz & / & 10 kHz & / \\
\hline
$\Delta E_{Z}$ & 103 MHz & 300 MHz & $\sim$100 MHz & 40 MHz & 46.9 MHz & 112 MHz \\
\hline
\end{tabular}
\label{Table_Expr}
\end{table*}

The robust control protocol proposed in this work is directly applicable to current-generation semiconductor spin qubit platforms. As summarized in Table~\ref{Table_Expr}, several experimental systems have already entered a regime where crosstalk noise becomes non-negligible, particularly when the ratio of drive strength to Zeeman energy splitting exceeds $\Omega/\Delta E_z > 0.5$. This condition enhances the visibility of parasitic terms such as $XZ$ and $YZ$ in the effective Hamiltonian, which our protocol is specifically designed to suppress.
For example, in the Ge/SiGe platforms reported by Hendrickx~\cite{Hendrickx2020N} and Wang~\cite{WangCA2024}, the effective drive strength reaches 25–40 MHz while the Zeeman splitting remains in the 40–50 MHz range, placing the system well into the high-drive regime where crosstalk becomes a limiting factor for gate fidelity. Similarly, for Si/SiGe systems such as those demonstrated by Xue~\cite{xue2022quantum}, Noiri~\cite{noiri2022fast}, and Philips~\cite{philips2022universal}, the drive strengths (~$5$ MHz) are already approaching a significant fraction of $\Delta E_z$, especially in systems with smaller Zeeman splittings. In this regime, control schemes that only compensate for static errors (e.g., $ZZ$ coupling) are insufficient, as performance rapidly degrades due to time-dependent crosstalk. Our general robust control pulses, which suppress both static and time-dependent noise channels, provide a crucial advantage for maintaining high gate fidelities under these experimentally relevant conditions.
Furthermore, since many platforms still operate with relatively low tunability in the exchange coupling $J$, often with limited $J_{\text{off}}$ or fixed $J$ architectures, the proposed protocol offers a practical and scalable route to high-fidelity quantum control without requiring additional hardware overhead.




\end{document}